\documentstyle[12pt]{article}

\catcode`\@=11
\long\def\@makefntext#1{
\protect\noindent \hbox to 3.2pt {\hskip-.9pt
$^{{\ninerm\@thefnmark}}$\hfil}#1\hfill}                %CAN BE USED

\def\@makefnmark{\hbox to 0pt{$^{\@thefnmark}$\hss}}  %ORIGINAL

\def\ps@myheadings{\let\@mkboth\@gobbletwo
\def\@oddhead{\hbox{}
\rightmark\hfil\ninerm\thepage}
\def\@oddfoot{}\def\@evenhead{\ninerm\thepage\hfil
\leftmark\hbox{}}\def\@evenfoot{}
\def\sectionmark##1{}\def\subsectionmark##1{}}

\renewcommand{\thefootnote}{\fnsymbol{footnote}}

\def\sectionc{\@startsection {section}{1}{\z@}{-3.5ex plus -1ex minus 
    -.2ex}{2.3ex plus .2ex}{\bf }}
\def\subsectionc{\@startsection{subsection}{2}{\z@}{-3.25ex plus -1ex minus 
   -.2ex}{1.5ex plus .2ex}{\it }}
\renewcommand{\section}[1]{\sectionc{#1}\hspace*{\parindent}}
\renewcommand{\subsection}[1]{\subsectionc{#1}\hspace*{\parindent}}
\newcounter{appendixc}
\newcounter{subappendixc}[appendixc]
\newcounter{subsubappendixc}[subappendixc]

\renewcommand{\appendix}[1] {\vspace*{0.6cm}
        \refstepcounter{appendixc}
        \setcounter{figure}{0}
        \setcounter{table}{0}
        \setcounter{equation}{0}
        \renewcommand{\thefigure}{\Alph{appendixc}.\arabic{figure}}
        \renewcommand{\thetable}{\Alph{appendixc}.\arabic{table}}
        \renewcommand{\theappendixc}{\Alph{appendixc}}
        \renewcommand{\theequation}{\Alph{appendixc}.\arabic{equation}}
        \noindent{\bf Appendix \theappendixc #1}\par\vspace*{0.4cm}}

\def\abstracts#1{{
        \centering{\begin{minipage}{13.2truecm}
        \footnotesize\baselineskip=13pt\noindent
        \parindent=0pt #1
        \end{minipage}}\par}}

\renewenvironment{thebibliography}[1]
        {\begin{list}{\arabic{enumi}.}
        {\usecounter{enumi}\setlength{\parsep}{0pt}
\setlength{\leftmargin 0.75cm}{\rightmargin 0pt}
         \setlength{\itemsep}{0pt} \settowidth
        {\labelwidth}{#1.}\sloppy}}{\end{list}}

\newcounter{itemlistc}
\newcounter{romanlistc}
\newcounter{alphlistc}
\newcounter{arabiclistc}

\newcommand{\fcaption}[1]{
        \refstepcounter{figure}
        \setbox\@tempboxa = \hbox{\footnotesize Figure~\thefigure. #1}
        \ifdim \wd\@tempboxa > 6in
           {\begin{center}
        \parbox{6in}{\footnotesize\baselineskip=13pt Figure~\thefigure. #1}
            \end{center}}
        \else
             {\begin{center}
             {\footnotesize Figure~\thefigure. #1}
              \end{center}}
        \fi}

\newcommand{\tcaption}[1]{
        \refstepcounter{table}
        \setbox\@tempboxa = \hbox{\footnotesize Table~\thetable. #1}
        \ifdim \wd\@tempboxa > 6in
           {\begin{center}
        \parbox{6in}{\footnotesize\baselineskip=13pt Table~\thetable. #1}
            \end{center}}
        \else
             {\begin{center}
             {\footnotesize Table~\thetable. #1}
              \end{center}}
        \fi}

\def\@citex[#1]#2{\if@filesw\immediate\write\@auxout
        {\string\citation{#2}}\fi
\def\@citea{}\@cite{\@for\@citeb:=#2\do
        {\@citea\def\@citea{,}\@ifundefined
        {b@\@citeb}{{\bf ?}\@warning
        {Citation `\@citeb' on page \thepage \space undefined}}
        {\csname b@\@citeb\endcsname}}}{#1}}

\newif\if@cghi
\def\cite{\@cghitrue\@ifnextchar [{\@tempswatrue
        \@citex}{\@tempswafalse\@citex[]}}
\def\citelow{\@cghifalse\@ifnextchar [{\@tempswatrue
        \@citex}{\@tempswafalse\@citex[]}}
\def\@cite#1#2{{$\null^{#1}$\if@tempswa\typeout
        {IJCGA warning: optional citation argument
        ignored: `#2'} \fi}}

 1
 1
 1

\font\ninerm=cmr9

\input{psfig}
\begin{document}
\hfill{\small {\bf MKPH-T-96-23}}\\
%\vspace*{0.6cm}
\begin{center}
\normalsize\bf New Insights into the Coupling of 
$\eta $ and $f_1$ Mesons to the Nucleon\footnote{Contribution to Int.\ Symp.\
Non-Nucleonic Degrees of Freedom Detected in Nucleus, Sept.\ 2-5, 1996
(Osaka, Japan)}
\end{center}
\baselineskip=15pt
%\vfill
\vspace*{0.6cm}
\centerline{\footnotesize M. KIRCHBACH$^1$, L. TIATOR$^1$,
 S. NEUMEIER$^2$, and S. KAMALOV$^3$}
\baselineskip=13pt
\centerline{\footnotesize\it $^1$Institut f\"ur Kernphysik, J.\ 
Gutenberg--Universit\"at, D--55099 Mainz, Germany}
\centerline{\footnotesize\it $^2$Institut f\"ur Kernphysik, TH Darmstadt, 
D--64289 Darmstadt, Germany}
\centerline{\footnotesize\it $^3$Departamento de Fisica Teorica,
  Universidad de Valencia, 46100 Burjassot, Spain}

%\vfill
\vspace*{0.6cm}
\abstracts{We show that the contact couplings of
neutral pseudoscalar and axial mesons to the  
nucleon are proportional to $\Delta s$, the fraction of nucleon spin carried
by the strange quark sea, and thus are strongly suppressed relative
to the couplings of charged mesons to the nucleon.
On the other side, recent high accuracy data on $\eta $ photoproduction at 
threshold
reveal the need for non--negligibile $\eta NN$ vertices, while fitting
$\bar N N$ phase shifts by means of effective meson exchange
potentials requires a substantial presence of $f_1$ mesons there. 
We here advocate the idea to attribute the couplings of neutral 
pseudoscalar and
axial vector mesons to effective triangular diagrams containig 
non--strange mesons
and demonstrate its usefulness in describing available data.  }

%\vspace*{0.6cm}
\normalsize\baselineskip=15pt
\setcounter{footnote}{0}
\renewcommand{\thefootnote}{\alph{footnote}}
\vspace*{0.6cm}
The neutral axial vector current of the nucleon
evaluated within the Glashow--Weinberg--Salam 
theory of electroweak interaction
is determined by the left--handed quark fields $\psi_L $ 
only in accordance with
\begin{eqnarray}
\langle N\mid J_{\mu ,5} ^ {neutral} \mid N\rangle 
= &&-2\langle N |\sum_{i=1}^{3} \bar \psi_L(i)\gamma_\mu {\tau_3^{weak}\over 2} 
\psi_L(i)
|N\rangle \nonumber\\
=&& - g_A \, \bar {\cal U}_N \gamma_\mu \gamma_5 
{\tau_3^{strong}\over 2} {\cal U}_N
+{{\Delta s}\over 2}\, \bar {\cal U}_N\gamma_\mu\gamma_5 {\cal U}_N
+ h.\, fl. \nonumber\\
= &&\langle N| \bar \Psi_q  \gamma_\mu \gamma_5 ({ \lambda_3\over 2}
-2 \sqrt{{1\over 3}}{\lambda_8\over 2} +\sqrt{{2\over 3 }}{\lambda_0\over 2} )
\Psi_q
|N\rangle \,  + h.\,  fl.
\label{Z0}
\end{eqnarray}
Here ${\tau_3^{weak}\over 2}/{\tau_3^{strong}\over 2} $ denotes the 
third component of 
{\em weak/strong \/} isospin, 
$i$ counts the quark generation, $\Psi_q$ stands for $col(u,d,s)$,
$\lambda_i$ are the standard Gell-Mann matrices,
while $\Delta s =S^\mu \langle N|\bar s\gamma_\mu \gamma_5 s|N\rangle $ is 
the fraction of nucleon spin carried by the strange quark sea.
The first quark generation acts simultaneously as a doublet with respect to
both weak and strong isospin. On the contrary,
each heavy flavor ($h.\, fl.$) quark generation decomposes
with respect to strong isospin into singlets.
As a consequence, $J_{\mu ,5}^{neutral}$
contains only the antisymmetric combination
$(\bar u u - \bar d d)/2$ which is identical to the third component 
of the charged axial current, while 
the symmetric combination $(\bar u u + \bar d d) $ is completely absent. 
This circumstance is of 
crucial importance for the design of the couplings of neutral $0^-$ and $1^+$
mesons to the nucleon. Indeed, in assuming the $(\bar q_j q_j)_M$
component of a neutral meson (M) to the neutral axial vector current 
$\bar q_k \gamma_\mu \gamma_5 q_k$ of quarks $q_k$ of flavor $k$ to be diagonal
 in flavor  and universal in strength\cite{Jaffe,KiRi}
\begin{equation}
 \langle \bar l l|\bar q_k \gamma_\mu \gamma_5 q_k|(\bar q_jq_j)_M\rangle =
\kappa \, \tilde{m}_M iq_\mu^M \delta_{kj}\, ,
\label{diag_fl}
\end{equation}
where $\tilde{m}_M$ is a properly chosen mass scale parameter and $\bar l l$ 
denotes the external leptons, it 
obviously occurs that the $\eta $ and $f_1$ mesons,
\begin{eqnarray}
\eta & = & {1\over \sqrt{6}} (\bar u u + \bar d d -2 \bar s s)\, ,\nonumber\\
f_1(1285) &=&  \cos\theta {{\bar u u + \bar d d }\over \sqrt{2}} 
- \sin\theta \bar s s\, , \nonumber\\
f_1(1420) &=& \sin\theta {{\bar u u + \bar d d }\over \sqrt{2}} + \cos \theta
\bar s s\, , \qquad \theta \approx 17^o\, ,
\label{mesons}
\end{eqnarray}
can couple to the current in Eq.~(\ref{Z0}) only through their 
$\bar s s$ components.
This because the non--strange pieces of the
octet ($J_{\mu ,5}^{(8)}$) and singlet ($J_{\mu ,5}^{(0)}$)
 axial vector currents 
\begin{eqnarray}
\langle N\mid J_{\mu ,5}^ {(8)}\mid N\rangle &=&
\langle N\mid \bar \Psi_q \gamma_\mu \gamma_5 {\lambda^8\over 2}
\Psi_q\mid N\rangle
= G_A^{(8)} \, \bar {\cal U}_N \gamma_\mu \gamma_5 {\cal U}_N\, , \nonumber\\
G_A^{(8)} &= &{{3F-D}\over {2\sqrt{3}}} = 
{1\over {2\sqrt{3}}} (\Delta u +\Delta d -2\Delta s)\, ,\nonumber\\
\langle N\mid J_{\mu ,5}^ {(0)}\mid N\rangle &=&
\langle N\mid \bar \Psi_q \gamma_\mu \gamma_5 {\lambda_0\over 2}
\Psi_q\mid N\rangle
= G_A^{(0)} \, \bar {\cal U}_N \gamma_\mu \gamma_5 {\cal U}_N\, ,\nonumber\\
G_A^{(0)} = \sqrt{{1\over 6}}(\Delta u +\Delta d +\Delta s) &, & 
-2\sqrt{{1\over 3}} G_A^{(8)} +\sqrt{{2\over 3}}G_A^{(0)} = \Delta s \, ,
\label{GA8}
\end{eqnarray}
cancel in the combination determining the neutral axial vector current
in Eq.~(\ref{Z0}), and the isosinglet part of $J_{\mu ,5}^{neutral}$ 
in fact coincides with the strange axial vector current
(in the approximation in which
heavier flavors have been omitted). 
For this reason both the $\eta NN$ and $f_1 NN$ vertices in fact 
appear proportional to
$\Delta s $ only and are thus, for $\Delta s = -0.13 \pm 0.04$\cite{Jaffe},
suppressed by a factor of $\Delta s/g_A \sim 0.1$ 
as compared to the coupling of the charged $\pi $ and $a_1$ mesons to the
nucleon\cite{KiTi}.
The value of the pseudoscalar $\eta NN$ coupling constant
$g_{\eta NN}$ predicted by the constituent quark model to satisfy the relation
$g_{\eta NN} = 2G_A^{(8)} g_{\pi NN} /g_A $
(with $g_A = \Delta u - \Delta d = 1.26 $)
appears therefore substantially 
overestimated as compared to the physical $\eta NN$ coupling
as $G_A^{(8)}$ contains in addition to $\Delta s$ also
the superfluous $\Delta u$ and $\Delta d$ contributions.
Regarding the $f_1$ mesons, the constituent quark model
even completely fails in predicting their couplings to the nucleon since these
particles behave as mesonic molecules $(\bar K K $ mesonium)
rather than as the $\bar q q$ states given in Eq.~(\ref{mesons}).

{}On the other side, recent high accuracy data on $\eta $ photoproduction 
at threshold\cite{Kru}
reveal the need for non--negligibile $\eta NN$ vertices,  while fitting
$\bar N N$ phase shifts by means of effective meson exchange
potentials requires a substantial presence of $f_1$ mesons 
there\cite{TiBeKa,Mull}. 
We here put forward the idea that the $\eta NN$ and $f_1(1285)NN$
couplings are mainly governed by the 
$a_0(980)\pi N$ triangle diagram, while the $f_1 (1420)NN $ coupling 
is associated with the
$\bar K^*(892) K (Y =\Sigma, \Lambda) $ effective triangular 
vertex\cite{KiTi,KiRi,Stefan}.
To illustrate this method we present analytical expressions
for the pseudoscalar as well as pseudovector (denoted by $f_{\eta NN}$)
coupling constants resulting from the evaluation of the $a_0\pi N$ triangle.
Details on the Lagrangians and the notations used are given
in previous work\cite{KiTi}.
The following expression for 
the $\eta NN$ vertex have been found: 
\begin{eqnarray}
g_{\eta NN}(q^2) & = &
\frac{3}{8\pi^2}{{m_{a_0}^2-m_\eta^2}\over m_\pi^2}
 f_{\pi NN}f_{a_0\eta \pi}g_{a_0 NN}\nonumber\\
& & \biggl\{   \int_0^1dx \ln
 \frac{{\cal Z}_1(m_\pi ,\Lambda_{a_0}, x,q^2)
{\cal Z}_1(\Lambda_\pi,m_{a_0}, x,q^2)}
{ {\cal Z}_1(m_\pi,m_{a_0}, x,q^2)
{\cal Z}_1(\Lambda_\pi,\Lambda_{a_0} , x,q^2)}   \nonumber\\
& +& \int_0^1\int_0^1 dy dx {{xc(x,y,1-y, q^2)}\over
{{\cal Z}_2(m_\pi,m_{a_0}, x,y, 1-y,q^2)}} \nonumber\\
&+ &  {1\over 2}\int_0^1\int_0^1 dy dx x \Bigl( \ln
{ { {\cal Z}_2 (m_\pi,\Lambda_{a_0}, x,y, 1-y,q^2) } \over
{ {\cal Z}_2(m_\pi,m_{a_0}, x,y, 1-y,q^2)} }\nonumber\\
& + &  \ln 
{ { {\cal Z}_2(\Lambda_\pi ,m_{a_0}, x,y,1-y,q^2)}\over
{ {\cal Z}_2(\Lambda_\pi,\Lambda_{a_0} , x,y, 1-y,q^2)} } \Bigr) \biggr\}\,.
\label{g_etann}
\end{eqnarray}
The functions ${\cal Z}_1(m_1,m_2,x,q^2)$,
${\cal Z}_2(m_1,m_2, x,y, \bar y,q^2)$ and $c(x,y,1-y,q^2)$ read
\begin{eqnarray}
c(x, y,\bar y, q^2) & =& x\bar y (1+x\bar y )m_N^2 +
xy( x(y + \bar y) + {1\over 2})q^2 \, ,\nonumber\\
{\cal Z}_1(m_1,m_2,x,q^2) & = & xm_1^2 +(m_2^2-q^2)(1-x) 
+ (1-x)^2q^2\, , \nonumber\\
{\cal Z}_2(m_1,m_2, x,y,\bar y , q^2) & =& 
m_N^2x^2\bar y^2 +m_1^2(1-x) + (m_2^2- q^2)xy \nonumber\\
&& + x^2y(y+\bar y)q^2\, .
\label{functions}
\end{eqnarray}
The corresponding expression for the pseudovector coupling is obtained as:
\begin{eqnarray}
f_{\eta NN}(q^2) & = &
\frac{3}{8\pi^2}{{m_{a_0}^2-m_\eta^2}\over m_\pi^2}
 f_{\pi NN}f_{a_0\eta \pi}g_{a_0 NN} m_Nm_\eta \nonumber\\
& & \int_0^1\int_0^1 dy dx 2 x^2 y  
Z^{-1}_2 (m_\pi,m_{a_0}, x,y, 1-y,q^2)\,.
\label{f_etann}
\end{eqnarray}
Here the coupling constant $f_{a_0\pi \eta}$ of the 
$a_0\to \pi + \eta $ decay has the value of $f_{a_0\eta \pi }=0.4$ 
corresponding to a decay width of $50$ MeV, while
$\Lambda_\pi$ and $\Lambda_{a_0}$ are the cut-off parameters
in the monopole vertex factors, for which we use 
the values 1.05 GeV and 2.0 GeV, respectively.
For the $a_0 (980) NN$ coupling constant, the smallest value of
$g^2_{a_0 NN}/4\pi = 0.77$ available from the
Bonn potential has been chosen. 
The $\eta NN$ coupling constants are obtained by setting 
$q^2= 0 $ in Eqs.~(\ref{g_etann}), (\ref{functions}) and (\ref{f_etann})
yielding 
\begin{eqnarray}
g_{\eta NN} (q^2 =0) & = &2.03\, ,\qquad 
{g_{\eta NN}^2\over {4\pi}} =0.33\, ,\\
f_{\eta NN} (q^2 =0) & = & 0.58\, , \qquad {f_{\eta NN}^2\over 
{4\pi}} = 0.03\, .
\label{values}
\end{eqnarray}
As an example for the power and usefulness of the
method suggested, we present in Fig. 1 the rather satisfactory description of
the high accuracy data\cite{Kru} on the
differential cross section for $\eta $ photoproduction off the proton
at a photon laboratory energy of 724 MeV. 
\begin{figure}[htbp]
\centerline{\psfig{figure=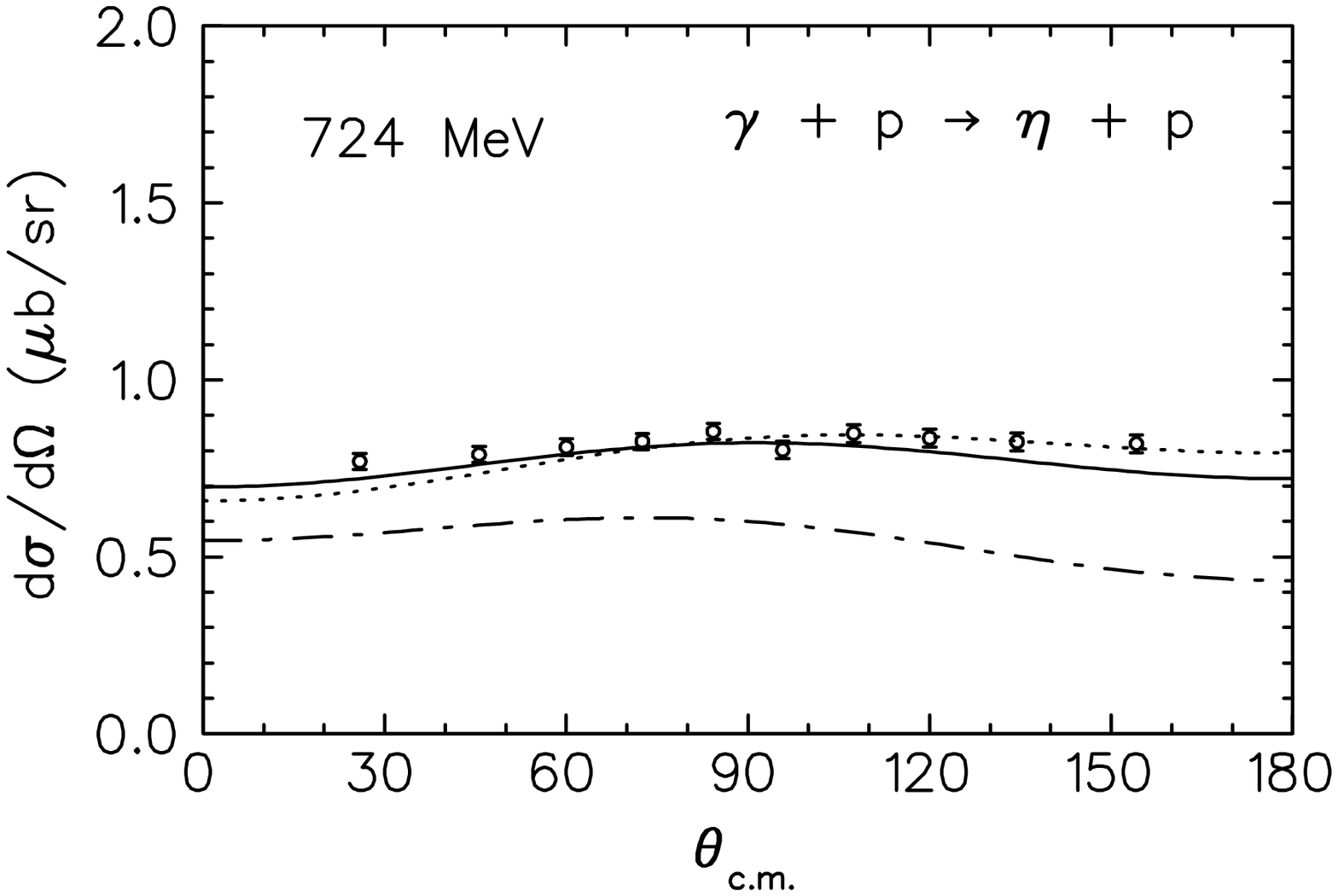,width=6cm}}
\vspace{0.0cm}
{\small 
Fig. 1 \hspace{0.2cm} Differential cross section for eta photoproduction
on the proton.} 
\end{figure}
In a similar way,  corresponding expressions have been found for
the $f_1(1285)NN$ (Ref.~\cite{KiRi}) and $f_1(1420)NN$ (Ref.~\cite{Stefan}) 
coupling constants (in turn denoted by $g_{f_1(1285)NN}$ and 
$g_{f_1(1420) NN}$) entering  vetrices
of the type    $g_{f_1 NN} \epsilon_{f_1}^\mu \bar 
{\cal U}_N \gamma_\mu \gamma_5
{\cal U}_N$,  respectively, with $\epsilon_{f_1}^\mu$ standing for the
polarisation vector of the $f_1$ meson.
With the parameters listed above, one finds
$g_{f_1(1285)NN} = 1.46$ for $q^2 = 27fm^{-2}$. 
A remarkable feature of the $f_1(1420)NN$ coupling, 
when associated with the $\bar K^*(892) K Y$ intermediate triangular states,
is that it appears predominantly induced by the tensor
coupling of the $K^*(892)$ meson to the baryon.
In using a value of 4.5 for the ratio between the tensor and vector couplings
of the $K^*(892)$ mesons in the $K^*(892)\Lambda N $ vertex,
a value of $g_{f_1(1420)NN} = 10.5$ has been reported in\cite{Stefan}
where also the Lagrangians used are presented.
A further interesting aspect of the $f_1 NN$ effective vertex discussed 
in\cite{Stefan} is that
the effective $\bar K^*(892) KY$ triangles induce a small tensor coupling
of the $f_1(1420)$ meson to the nucleon, which is, however, by about two order
of magnitudes smaller as compared to $g_{f_1(1420)NN}$.
As it is well known, such a coupling is G-parity forbidden.
So, second class currents may be induced by the effective vertices considerded
above. 
To conclude, we wish to stress that the couplings of the mesons occupying the
centers of the $0^-$ and $1^+$  meson nonets
cannot be described in the context of the constituent quark model
which does not account for the structure of the neutral
axial vector current emerging in the gauge theory of electroweak
interaction of the current quarks.


\begin{thebibliography}{9}
\bibitem{Jaffe} R.\ L.\ Jaffe, Phys.\ Lett.\ {\bf B229} (1989) 275
\bibitem{KiRi} M.\ Kirchbach and D.\ O.\ Riska, Nucl.\ Phys.\ 
{\bf A594} (1995) 419
\bibitem{KiTi}  M.\ Kirchbach and L.\ Tiator, Nucl.\ Phys.\ 
{\bf A604} (1996) 385
\bibitem{Kru} B.\ Krusche et al.\ , Phys.\ Rev.\ Lett.\ {\bf 74} (1995) 3736
\bibitem{TiBeKa} L.\ Tiator, C.\ Bennhold and S.\ S.\ Kamalov, 
                 Nucl.\ Phys.\ {\bf A580} (1994) 455
\bibitem{Mull} V.\ Mull, Ph.\ D.\ thesis, University Bonn, 1993, Germany
\bibitem{Stefan} S.\ Neumeier, diploma thesis, TH Darmstadt, 1996, Germany  
\end{thebibliography}
\end{document}